\documentclass[aps,prd,twocolumns,eqsecnum,preprintnumbers,showpacs,amsmath,amssymb]
{revtex4}

\usepackage{graphicx}
\usepackage{dcolumn}
\usepackage{bm}

\begin{document}

\title{THE COLOR GAUGE INVARIANCE  \\ AND A POSSIBLE ORIGIN OF A MASS GAP IN QCD }

\author{V. Gogokhia}
\email[]{gogohia@rmki.kfki.hu}

\affiliation{HAS, CRIP, RMKI, Depart. Theor. Phys., Budapest 114,
P.O.B. 49, H-1525, Hungary}

\date{\today}
\begin{abstract}
The general scale parameter, having the dimensions of mass
squared, is dynamically generated in the QCD gluon sector. It is
introduced through the difference between the regularized full
gluon self-energy and its value at some finite point. It violates
transversality of the full gluon self-energy. The Slavnov-Taylor
identity for the full gluon propagator, when it is given by the
corresponding equation of motion, is also violated by it. So in
order to maintain both transversality and the identity it should
be disregarded from the very beginning, i.e., put formally zero
everywhere. However, we have shown how to preserve the
above-mentioned identity at non-zero mass squared parameter. This
allows one to establish the structure of the full gluon propagator
when it is explicitly present. Its contribution does not survive
in the perturbation theory regime, when the gluon momentum goes to
infinity. At the same time, its contribution dominates the
structure of the full gluon propagator when the gluon momentum
goes to zero. We have also proposed a method how to restore
transversality of the relevant gluon propagator in a gauge
invariant way, while keeping the mass squared parameter "alive".

\end{abstract}



\maketitle

\section{Introduction}

Quantum Chromodynamics (QCD) \cite{1,2,3,4} is widely accepted as
a realistic quantum field gauge theory of the strong interactions
not only at the fundamental (microscopic) quark-gluon level but at
the hadronic (macroscopic) level as well. It is a $SU(3)$ color
gauge invariant theory but:

{\bf (i)}. Due to color confinement, the gluon (unlike the photon)
is not a physical state. Moreover, there is no physical amplitude
to which the gluon self-energy (like the photon self-energy) may
directly contribute.

{\bf (ii)}. In contrast to the conserved currents in Quantum
Electrodynamics (QED), the color-conserved currents do not play
any role in the extraction of physical information from the
$S$-matrix elements for the corresponding physical processes and
quantities in QCD. In other words, the conserved color currents do
not contribute directly to the $S$-matrix elements describing this
or that physical process/quantity. For this their color-singlet
counterparts, which can even be partially conserved, are relevant.
For example, an important physical QCD parameter such as the pion
decay constant is given by the following $S$-matrix element:
$<0|J^i_{5\mu}(0)|\pi^j(q)>= i q_{\mu} F_{\pi} \delta^{ij}$, where
$J^i_{5\mu}(0)$ is the axial-vector current, while $|\pi^j(q)>$
describes the pion bound-state amplitude, and $i, j$ are flavor
indices.

{\bf (iii)}. In QCD (contrary to QED) there exists direct
evidence/indication that transversality of the full gluon
self-energy, as well as the Slavnov-Taylor (ST) identity for the
full gluon propagator, as it is determined by the corresponding
equation of motion, is violated. Indeed, there is no
regularization scheme (preserving or not gauge invariance) in
which the transversality condition and the ST identity could be
satisfied unless the so-called constant skeleton tadpole term (or,
equivalently, its re-defined counterpart, which we will call the
general mass scale parameter in what follows) is to be disregarded
from the very beginning, i.e., put formally zero everywhere.

However, our main goal in this paper is to show that the tadpole
term is consistent with the color gauge invariance in QCD, i.e.,
it is maintained at non-zero general mass scale parameter as well.

\section{The full gluon self-energy}

For our purpose it is convenient to begin with the general
description of the Schwinger-Dyson (SD) equation for the full
gluon propagator. It can be written as follows:

\begin{equation}
D_{\mu\nu}(q) = D^0_{\mu\nu}(q) + D^0_{\mu\rho}(q) i
\Pi_{\rho\sigma}(q; D) D_{\sigma\nu}(q),
\end{equation}
where

\begin{equation}
D^0_{\mu\nu}(q) = i \left\{ T_{\mu\nu}(q) + \xi L_{\mu\nu}(q)
\right\} {1 \over q^2}
\end{equation}
is the free gluon propagator, and $\xi$ is the gauge-fixing
parameter. Also, here and everywhere below $T_{\mu\nu}(q) =
\delta_{\mu\nu} - (q_{\mu} q_{\nu} / q^2) =  \delta_{\mu\nu} -
L_{\mu\nu}(q)$, as usual in Euclidean metrics (see remarks below
as well). $\Pi_{\rho\sigma}(q; D)$ is the full gluon self-energy
which depends on the full gluon propagator due to the non-abelian
character of QCD. Thus the gluon SD equation is highly nonlinear
(NL). Evidently, we omit the color group indices, since for the
gluon propagator (and hence for its self-energy) they factorize,
for example $D^{ab}_{\mu\nu}(q) = D_{\mu\nu}(q)\delta^{ab}$.

\begin{figure}[h]
\begin{center}
\includegraphics[width=0.8\textwidth]{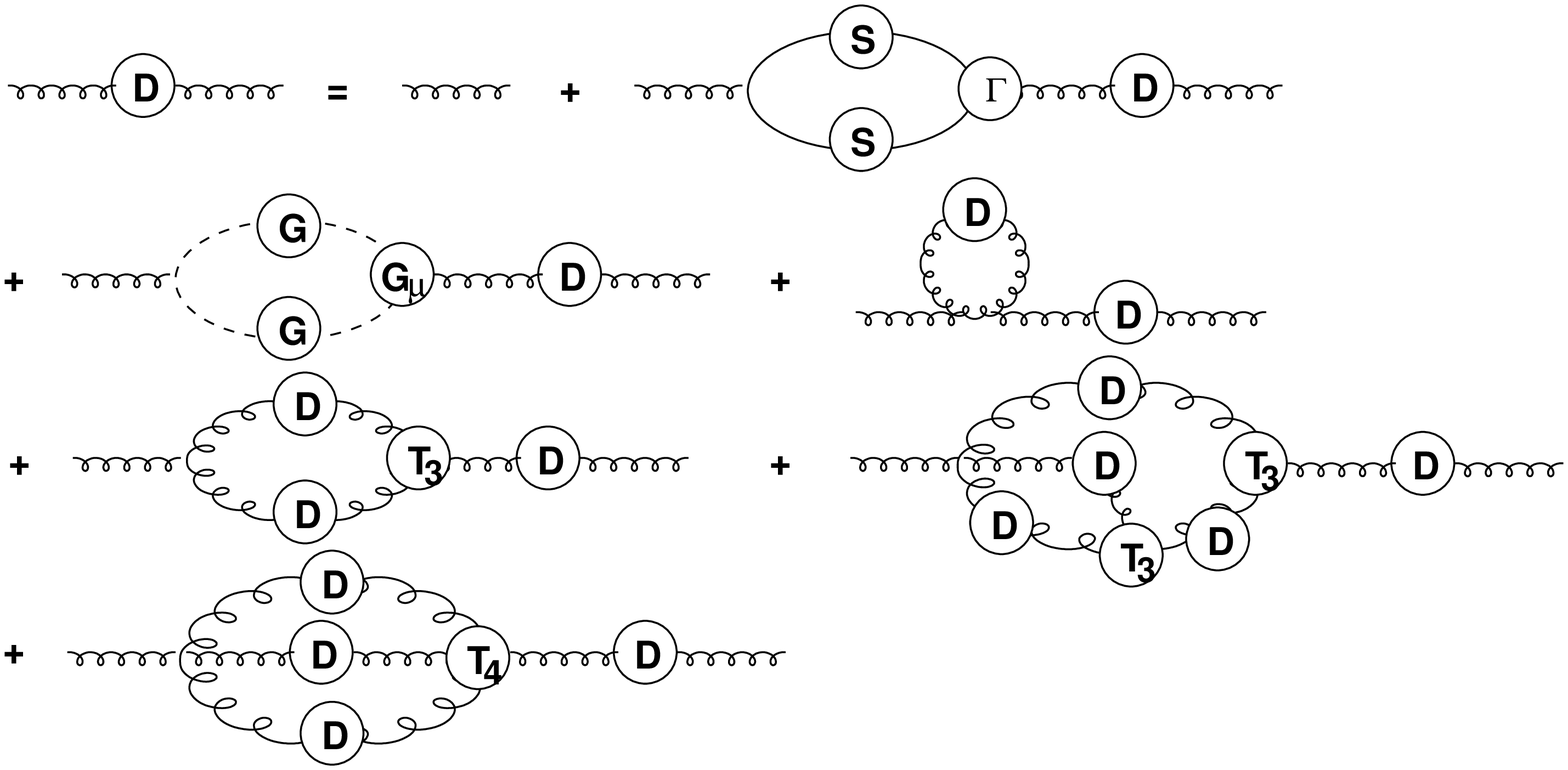}
\caption{The SD equation for the full gluon propagator. }
\label{fig:1}
\end{center}
\end{figure}

The full gluon self-energy $\Pi_{\rho\sigma}(q; D)$ is the sum of
a few terms (see Fig. 1),

\begin{equation}
\Pi_{\rho\sigma}(q; D)=  \Pi^q_{\rho\sigma}(q) +
\Pi^{gh}_{\rho\sigma}(q) + \Pi_{\rho\sigma}^t(D) +
\Pi^{(1)}_{\rho\sigma}(q; D^2) + \Pi^{(2)}_{\rho\sigma}(q; D^4) +
\Pi^{(2')}_{\rho\sigma}(q; D^3),
\end{equation}
where $\Pi^q_{\rho\sigma}(q)$ describes the skeleton loop
contribution due to the quark degrees of freedom (it is an analog
of the vacuum polarization tensor in QED), while
$\Pi^{gh}_{\rho\sigma}(q)$ describes the skeleton loop
contribution associated with the ghost degrees of freedom. Since
neither of the skeleton loop integrals depends on the full gluon
propagator $D$, they represent the linear contribution to the
gluon SD equation. $\Pi_{\rho\sigma}^t(D)$ is the so-called
constant skeleton tadpole term. $\Pi^{(1)}_{\rho\sigma}(q; D^2)$
represents the skeleton loop contribution, which contains the
triple gluon vertices only. $\Pi^{(2)}_{\rho\sigma}(q; D^4)$ and
$\Pi^{(2')}_{\rho\sigma}(q; D^3)$ describe topologically
independent skeleton two-loop contributions, which combine the
triple and quartic gluon vertices. All these quantities are given
by the corresponding loop diagrams in Fig. 1. The last four terms
explicitly contain the full gluon propagators in the corresponding
powers symbolically shown above. They thus form the NL part of the
gluon SD equation. The analytical expressions for the
corresponding skeleton loop integrals \cite{5} (in which the
symmetry coefficients and signs have been included, for
convenience) are of no importance here, since we are not going to
introduce into them any truncation/approximation or choose some
special gauge. Let us note in advance that here and below the
signature is Euclidean, since it implies $q_i \rightarrow 0$ when
$q^2 \rightarrow 0$ and vice-versa. All the quantities which
contribute to the full gluon self-energy (2.3) are tensors, having
the dimensions of mass squared. All these skeleton loop integrals
are therefore quadratically divergent in perturbation theory (PT),
and so they are assumed to be regularized, as discussed below.

\section{The Subtractions}

Let us subtract from the full gluon self-energy (2.3) its value at
$q=0$. Thus, one obtains

\begin{equation}
\Pi^s_{\rho\sigma}(q; D) = \Pi_{\rho\sigma}(q; D) -
\Pi_{\rho\sigma}(0; D) = \Pi_{\rho\sigma}(q; D) -
\delta_{\rho\sigma} \Delta^2 (D),
\end{equation}
which is nothing but the definition of the subtracted full gluon
self-energy $\Pi^s_{\rho\sigma}(q; D)$. Contrary to QED, QCD being
a non-abelian gauge theory can suffer from infrared (IR)
singularities in the $q^2 \rightarrow 0$ limit due to the
self-interaction of massless gluon modes. Thus the initial
subtraction at zero in the definition (3.1) may be dangerous
\cite{1}. That is why in all the quantities below the dependence
on the finite (slightly different from zero) dimensionless
subtraction point $\alpha$ is to be understood. In other words,
all the subtractions at zero and the Taylor expansions around zero
should be understood as the subtractions at $\alpha$ and the
Taylor expansions near $\alpha$, where they are justified to be
used. From a technical point of view, however, it is convenient to
put formally $\alpha=0$ in all the derivations below, and to
restore the explicit dependence on non-zero $\alpha$ in all the
quantities only at the final stage. At the same time, in all the
quantities where the dependence on $\lambda$ (which is the
dimensionless ultraviolet (UV) regulating parameter) and $\alpha$
is not shown explicitly, nevertheless, it should be assumed. For
example, $\Delta^2(D) \equiv \Delta^2(\lambda, \alpha; D)$ and
similarly for all other quantities (evidently, in all the
subtracted quantities like this one $D$ depends on the loop(s)
variable(s) only). This means that all the expressions are
regularized (i.e., they become finite), and thus a mathematical
meaning is assigned to all of them. For our purpose, in principle,
it is not important how $\lambda$ and $\alpha$ have been
introduced. They should be removed at the final stage only as a
result of the self-consistent renormalization program.

From the subtraction (3.1) it follows that the general scale
parameter $\Delta^2 (D)$, having the dimensions of mass squared,
is dynamically generated in the QCD gluon sector. It is defined as
the value of the regularized full gluon self-energy at some finite
point (see discussion above). It is mainly due to the nonlinear
interaction of massless gluon modes plus the linear contributions
from quark and ghost degrees of freedom, namely

\begin{equation}
\Delta^2 (D)= \Pi_t(D) + \Pi_q(0)+ \Pi_g(0; D) = \Delta^2_t(D) +
\Delta^2_q + \Delta^2_g(D),
\end{equation}
where

\begin{equation}
\Delta^2_g (D) \equiv \Pi_g(0; D) = \sum_a \Pi_a(0; D) = \sum_a
\Delta^2_a(D),
\end{equation}
and the index "a" runs as follows: $a= gh, (1), (2), (2')$. The
tensor indices have been omitted, so in this case all the indices
$t, q, g$, and hence $a$, are subscripts. In these relations all
the quadratically divergent constants $\Pi_t(D) \equiv
\Delta^2_t(D)$, $ \Pi_q(0) \equiv \Delta^2_q$, and $\Pi_a(0; D)
\equiv \Delta^2_a(D)$, having the dimensions of mass squared, are
given by the corresponding regularized skeleton loop integrals at
$q^2=0$ that appear in Eq.~(2.3). Let us remind that no
truncations/approximations/assumptions, and no special gauge
choice are made in the above-mentioned loop integrals, which
contribute to the regularized full gluon self-energy.

The subtracted gluon self-energy (3.1)

\begin{equation}
\Pi^s_{\rho\sigma}(q; D) \equiv \Pi^s(q; D) = \Pi^s_q(q) +
\Pi^s_g(q; D) = \Pi^s_q(q) + \sum_a \Pi^s_a(q; D)
\end{equation}
is free of the tadpole contribution, because $\Pi^s_t(D) =
\Pi_t(D)- \Pi_t(D)=0$, by definition, at any $D$, while in the
gluon self-energy (2.3) it is explicitly present.

\section{Transversality of the full gluon self-energy}

Contracting the full gluon self-energy (2.3) with $q_{\rho}$, it
can be reduced to the two independent transversality conditions,
namely

\begin{equation}
q_{\rho} \Pi_{\rho\sigma}(q; D)= q_{\rho} \Pi^q_{\rho\sigma}(q) +
q_{\rho} \Pi^g_{\rho\sigma}(q; D),
\end{equation}
where the pure gluon contribution is defined as follows:

\begin{equation}
\Pi^g_{\rho\sigma}(q; D)= \Pi_{\rho\sigma}^t(D) +
\Pi^{gh}_{\rho\sigma}(q) + \Pi^{(1)}_{\rho\sigma}(q; D^2) +
\Pi^{(2)}_{\rho\sigma}(q; D^4) + \Pi^{(2')}_{\rho\sigma}(q; D^3).
\end{equation}

\subsection{The quark contribution}

The color currents conservation condition (quite similar to the
current conservation in QED) implies

\begin{equation}
q_{\rho} \Pi^q_{\rho\sigma}(q)= q_{\sigma} \Pi^q_{\rho\sigma}(q) =
0.
\end{equation}
Let us show that then the corresponding constant skeleton quark
loop contribution $\Delta^2_q$ to Eq. (3.2) has to be discarded
from the very beginning. For this purpose, it is instructive to
show the subtraction of the quark part explicitly as follows:

\begin{equation}
\Pi^{s(q)}_{\rho\sigma}(q) = \Pi^q_{\rho\sigma}(q) -
\Pi^q_{\rho\sigma}(0) = \Pi^q_{\rho\sigma}(q) -
\delta_{\rho\sigma} \Delta^2_q.
\end{equation}
At the same time, the general decompositions of the quark part and
its subtracted counterpart into the independent tensor structures
are

\begin{eqnarray}
\Pi^q_{\rho\sigma}(q) &=& T_{\rho\sigma}(q) q^2 \Pi(q^2) +
q_{\rho} q_{\sigma} \tilde{\Pi}(q^2), \nonumber\\
\Pi^{s(q)}_{\rho\sigma}(q) &=& T_{\rho\sigma}(q) q^2 \Pi^s(q^2) +
q_{\rho} q_{\sigma} \tilde{\Pi}^s(q^2).
\end{eqnarray}
Here and everywhere below all the invariant functions are
dimensionless ones of their argument $q^2$. In addition both
invariant functions $\Pi^s(q^2)$ and $\tilde{\Pi}^s(q^2)$ cannot
have the power-type singularities (or, equivalently, the pole-type
ones) at small $q^2$, since $\Pi^{s(q)}_{\rho\sigma}(0) =0$, by
definition; see Eq.~(4.4); otherwise they remain arbitrary. Let us
also note that from the color currents conservation condition
(4.3) it follows that $\tilde{\Pi}(q^2) = 0$, so
$\Pi^q_{\rho\sigma}(q)$ is purely transversal, indeed.

On the other hand, contracting the relation (4.4) with $q_{\rho}$,
on account of the relation (4.3) and the second of the relations
(4.5), one obtains

\begin{equation}
\tilde{\Pi}^s(q^2) = - { \Delta^2_q \over q^2}.
\end{equation}
This, however, is impossible since $\tilde{\Pi}^s(q^2)$ cannot
have the power-type singularities at small $q^2$, as underlined
above. The only solution to the previous relation is to disregard
$\Delta^2_q$ from the very beginning, i.e., put formally zero, and
hence $\tilde{\Pi}^s(q^2)=0$ as well. Thus, one has

\begin{equation}
\Delta^2_q =0,
\end{equation}
so the subtracted quark part $\Pi^{s(q)}_{\rho\sigma}(q)$ becomes
also transversal $q_{\rho} \Pi^{s(q)}_{\rho\sigma}(q)=0$, and
$\Pi^q_{\rho\sigma}(q) = \Pi^{s(q)}_{\rho\sigma}(q)$, which yields
$\Pi(q^2) = \Pi^s(q^2)$. This describes a general situation, when
just the initial transversality condition (4.3) for
$\Pi^q_{\rho\sigma}(q)$ lowers the quadratic divergence of the
corresponding loop integral(s) to a logarithmic one at large
$q^2$. They may still present in its subtracted counterpart
$\Pi^{s(q)}_{\rho\sigma}(q)$. This is in complete analogy with QED
(see appendix A), since there only electron-positron skeleton loop
(the vacuum polarization tensor) contributes to the full photon
self-energy.

Let us underline that in obtaining this result no any
regularization scheme (preserving or not gauge invariance) has
been used. It has been also obtained without using PT (all the
full Green's functions which are present in the quark skeleton
loop integral have not been replaced by their free PT
counterparts). No special gauge choice has been made as well.

\subsection{The pure gluon contribution}

In the same way, however,

\begin{equation}
q_{\rho} \Pi^g_{\rho\sigma}(q; D)  = q_{\rho} \Bigl[
\Pi^t_{\rho\sigma}(D) + \Pi^{gh}_{\rho\sigma}(q) +
\Pi^{(1)}_{\rho\sigma}(q; D^2) + \Pi^{(2)}_{\rho\sigma}(q; D^4) +
\Pi^{(2')}_{\rho\sigma}(q; D^3) \Bigr] \neq 0,
\end{equation}
and hence $q_{\rho} \Pi_{\rho\sigma}(q; D) \neq 0$, unless the
constant skeleton tadpole term $\Pi^t_{\rho\sigma}(D)$ is
discarded from the very beginning. So omitting it in the relation
(4.8), one obtains

\begin{equation}
q_{\rho} \Bigl[ \Pi^{gh}_{\rho\sigma}(q) +
\Pi^{(1)}_{\rho\sigma}(q; D^2) + \Pi^{(2)}_{\rho\sigma}(q; D^4) +
\Pi^{(2')}_{\rho\sigma}(q; D^3) \Bigr] = 0.
\end{equation}
It should be noted that none of these quantities can satisfy the
corresponding transversality condition separately from each other,
i.e, similarly to the relation (4.3). The role of ghost degrees of
freedom is to cancel the unphysical (longitudinal) component of
the full gluon propagator. Therefore the transversality condition
(4.9) is important for ghosts to fulfill their role, and thus to
maintain unitarity of the $S$-matrix in QCD.

On account of the above-mentioned relation (4.7), the general
scale parameter (3.2) becomes

\begin{equation}
\Delta^2 (D)= \Delta^2_t(D) + \Delta^2_g(D).
\end{equation}
It is worth emphasizing that it does not matter whether we will
discard $\Delta^2_g(D)$ from the very beginning, in accordance
with the transversality condition (4.9) (in complete analogy with
the previous quark case), or not. In the explicit presence of the
tadpole term $\Delta^2_t(D)$ this is not important. Not losing
generality, it can be included into the tadpole term itself, and
the general mass scale parameter $\Delta^2 (D)$ in the relation
(4.10) is to be considered as the re-defined tadpole term.

Let us continue with the general decompositions of the full gluon
self-energy and its subtracted counterpart, which enter the
subtraction (3.1), as follows:

\begin{eqnarray}
\Pi_{\rho\sigma}(q; D) &=&  T_{\rho\sigma}(q) q^2 \Pi(q^2; D) +
q_{\rho} q_{\sigma} \tilde{\Pi}(q^2; D), \nonumber\\
\Pi^s_{\rho\sigma}(q; D) &=&  T_{\rho\sigma}(q) q^2 \Pi^s(q^2; D)
+ q_{\rho} q_{\sigma} \tilde{\Pi}^s(q^2; D),
\end{eqnarray}
where again all the invariant functions of $q^2$ are dimensionless
ones, while in addition the invariant functions $\Pi^s(q^2; D)$
and $\tilde{\Pi}^s(q^2; D)$ cannot have the pole-type
singularities in the $q^2 \rightarrow 0$ limit, since
$\Pi^s_{\rho\sigma}(0; D) =0$, by definition; otherwise all the
invariant functions remain arbitrary. Of course, they are
different from those of the relations (4.5).

Contracting them with $q_{\rho}$ along with the subtraction (3.1),
one obtains

\begin{equation}
\tilde{\Pi}(q^2; D) = \tilde{\Pi}^s(q^2; D) + {\Delta^2(D) \over
q^2},
\end{equation}
and

\begin{equation}
\Pi(q^2; D) = \Pi^s(q^2; D) + {\Delta^2(D) \over q^2}.
\end{equation}

It is worth emphasizing that the full gluon self-energy has a
massless single particle singularity due to non-zero mass scale
parameter $\Delta^2(D)$, which is of the non-pertubative (NP)
origin. At the same time, its subtracted counterpart cannot have
such a singularity, as mentioned above. In other words, this means
that in the explicit presence of the mass scale parameter both
invariant functions of the full gluon self-energy gain additional
contributions due to it (of course, not only at some finite
subtraction point $q^2=\mu^2 \neq 0$). If $\Delta^2(D)$ is welcome
in the transversal invariant function $\Pi(q^2; D)$, it is not
welcome in its longitudinal counterpart $\tilde{\Pi}(q^2; D)$,
since just it violates the ST identity. Let us note that
transversality of the full gluon self-energy together with its
subtracted counterpart can be achieved only in the formal
$\Delta^2(D)=0$ limit (for detail discussion of these preliminary
remarks see sections below). So in the general case of non-zero
$\Delta^2(D)$ only two possibilities remain.

(i). Both are not transversal and then

\begin{eqnarray}
q_{\rho} \Pi_{\rho\sigma}(q; D) &=& q_{\sigma} q^2
\tilde{\Pi}(q^2; D) = q_{\sigma} [q^2 \tilde{\Pi}^s(q^2; D)
+ \Delta^2(D)] \neq 0, \nonumber\\
q_{\rho} \Pi^s_{\rho\sigma}(q; D) &=& q_{\sigma} q^2
\tilde{\Pi}^s(q^2; D) = q_{\sigma} [q^2 \tilde{\Pi}(q^2; D) -
\Delta^2(D)].
\end{eqnarray}
The last inequality in the first of the relations (4.14) follows
from the fact that $\tilde{\Pi}^s(q^2; D)$ cannot have a single
particle singularity $ - \Delta^2(D)/ q^2$ in order to cancel
$\Delta^2(D)$.

(ii). Transversality of the subtracted gluon self-energy is
maintained, i.e., $\tilde{\Pi}^s(q^2; D)=0$ and then

\begin{equation}
q_{\rho} \Pi_{\rho\sigma}(q; D) = q_{\sigma} q^2 \tilde{\Pi}(q^2;
D) = q_{\sigma} \Delta^2(D) \neq 0, \quad q_{\rho}
\Pi^s_{\rho\sigma}(q; D) = 0,
\end{equation}
Contrary to the first case, now we know how precisely the
transversality of the full gluon self-energy is violated. So it is
always violated at non-zero mass scale parameter $\Delta^2(D)$. In
this connection one thing should be made perfectly clear. It is
the initial subtraction (3.1) which leaves the subtracted
gluon-self energy logarithmical divergent only, and hence the
invariant function $\Pi^s(q^2; D)$ is free of the quadratic
divergences, but a logarithmic ones can be still present in it, at
any $D$. Since the transversality condition for the full gluon
self-energy is violated in these relations, that is why we cannot
disregard $\Delta^2(D)$ from the very beginning (compare with the
pure quark case considered above).

\section{The ST identity for the full gluon propagator}

In order to calculate the physical observables in QCD from first
principles, we need the full gluon propagator rather than the full
gluon self-energy. The basic relation to which the full gluon
propagator should satisfy is the corresponding ST identity

\begin{equation}
q_{\mu}q_{\nu} D_{\mu\nu}(q) = i \xi.
\end{equation}
It is a consequence of the color gauge invarince/symmetry of QCD,
and therefore "is an exact constraint on any solution to QCD"
\cite{1}. This is true for any other ST identities. Being a result
of this exact symmetry, it is the general one, and it is important
for the renormalization of QCD. If some equation, relation or the
regularization scheme, etc.  do not satisfy it automatically,
i.e., without any additional conditions, then they should be
modified and not this identity (identity is an equality, where
both sides are the same, i.e., there is no room for additional
conditions). In other words, all the relations, equations,
regularization schemes, etc. should be adjusted to it and not vice
versa. It implies that the general tensor decomposition of the
full gluon propagator is

\begin{equation}
D_{\mu\nu}(q) = i \left\{ T_{\mu\nu}(q) d(q^2) + \xi L_{\mu\nu}(q)
\right\} {1 \over q^2},
\end{equation}
where the invariant function $d(q^2)$ is the corresponding Lorentz
structure of the full gluon propagator (sometimes we will call it
as the full effective charge ("running"), for simplicity). Let us
emphasize once more that these basic relations are to be satisfied
in any case, for example, whether the tadpole term itself or any
other mass scale parameter is put formally zero or not.

On account of the exact relations (4.11), (4.12) and (4.13), the
initial gluon SD equation (2.1) can be equivalently re-written
down as follows:

\begin{equation}
D_{\mu\nu}(q) = D^0_{\mu\nu}(q) + D^0_{\mu\rho}(q)i
T_{\rho\sigma}(q) [q^2 \Pi^s(q^2; D) + \Delta^2(D)]
D_{\sigma\nu}(q) + D^0_{\mu\rho}(q)i L_{\rho\sigma}(q) q^2
\tilde{\Pi}(q^2; D) D_{\sigma\nu}(q).
\end{equation}

Contracting this equation with $q_{\mu}$ and $q_{\nu}$, one
arrives at

\begin{equation}
q_{\mu}q_{\nu} D_{\mu\nu}(q) = i \xi - i \xi^2 \tilde{\Pi}(q^2;
D)= i \xi \left( 1 - \xi \tilde{\Pi}(q^2; D) \right),
\end{equation}
so the ST identity (5.1) is not automatically satisfied. In order
to get from this relation the ST identity, one needs to put

\begin{equation}
\tilde{\Pi}(q^2; D) =0,
\end{equation}
which is equivalent to

\begin{equation}
\tilde{\Pi}^s(q^2; D) = - { \Delta^2(D) \over q^2},
\end{equation}
as it follows from the relation (4.12). This, however, is
impossible since $\tilde{\Pi}^s(q^2; D)$ cannot have the
power-type singularities at small $q^2$, as underlined above. The
only solution to the previous relation is to disregard
$\Delta^2(D)$ from the very beginning, i.e., put formally zero

\begin{equation}
\Delta^2(D) =0
\end{equation}
everywhere, in complete analogy with the quark constant (4.7)
considered above. In this case from all the relations it follows
that the gluon full self-energy coincides with its subtracted
counterpart, and both quantities become purely transversal, since

\begin{equation}
\Pi(q^2; D^{PT}) = \Pi^s(q^2; D^{PT}), \quad \tilde{\Pi}(q^2;
D^{PT}) = \tilde{\Pi}^s(q^2; D^{PT})=0.
\end{equation}

The one way to satisfy the ST identity and thus to maintain the
color gauge structure of QCD is to discard the mass scale
parameter $\Delta^2(D)$ from the very beginning, i.e., put it
formally zero $\Delta^2(D)=0$ in all the equations, relations,
etc. In this limit the initial gluon SD equation (5.3) is modified
as follows:

\begin{equation}
D^{PT}_{\mu\nu}(q) = D^0_{\mu\nu}(q) + D^0_{\mu\rho}(q)i
T_{\rho\sigma}(q) q^2 \Pi^s(q^2; D^{PT})D^{PT}_{\sigma\nu}(q),
\end{equation}
and the corresponding Lorentz structure which appears in Eq. (5.2)
becomes

\begin{equation}
d^{PT}(q^2) = { 1 \over 1 + \Pi^s(q^2; D^{PT})}.
\end{equation}
It is easy to see that the gluon SD equation (5.9) automatically
satisfies the ST identity (5.1) now. Evidently, in the formal
$\Delta^2(D)=0$ limit we denote $D_{\mu\nu}(q)$ and $d(q^2)$ as
$D^{PT}_{\mu\nu}(q)$ and $d^{PT}(q^2)$, respectively (for reason
see section VII).

As it follows from above, in the formal $\Delta^2(D)=0$ limit the
full gluon self-energy and its subtracted counterpart (4.11)
become purely transversal, i.e.,

\begin{equation}
q_{\rho} \Pi_{\rho\sigma}(q; D^{PT}) = q_{\rho}
\Pi^s_{\rho\sigma}(q; D^{PT}) = 0.
\end{equation}
Let us remind that the invariant function $\Pi^s(q^2; D^{PT})$ can
be only logarithmical divergent in the $q^2 \rightarrow \infty$
limit. However, the formal $\Delta^2(D)=0$ limit implies
$\Delta^2_t(D)= \Delta^2_g(D)=0$, since they are independent
quantities, see Eq. (4.10) and remarks after it. As we already
know, in this case there will be no problems for ghosts to
accomplish their role, namely to cancel the longitudinal component
in the full gluon propagator (5.9). For the explicit cancellation
in lower order of PT, see, for example, Refs. \cite{2,3,4}.
However, such a cancellation should occur in every order of PT, in
accordance with the transversality condition (4.9).

We have reminded some important aspects of the color gauge
structure of QCD, but without any use of PT.

\subsection{A preliminary discussion}

A few clarifying remarks are in order already at this stage. By
substituting the initial gluon SD equation (2.1) into the ST
identity, on account of the first of the general decompositions
(4.11), one finally arrives at the conclusion that the full gluon
self-energy should be transversal, as it follows from the
condition (5.5). However, this is equivalent to the relation
(5.6), which has only unique solution (5.7), namely put formally
$\Delta^2(D)$ zero everywhere. The full gluon self-energy (4.11)
then becomes

\begin{equation}
\Pi_{\rho\sigma}(q; D^{PT}) =  T_{\rho\sigma}(q) q^2 \Pi(q^2;
D^{PT}) = T_{\rho\sigma}(q) q^2 \Pi^s(q^2; D^{PT}),
\end{equation}
due to the relations (5.5), (5.7), (5.8) and (4.13). This means
that if the gluon self-energy is transversal, then it cannot have
the pole-type singularity, since $\Pi(q^2; D^{PT}) = \Pi^s(q^2;
D^{PT})$, and $\Pi^s(q^2; D^{PT})$ is always regular at $q^2
\rightarrow 0$. In other words, when $\Delta^2(D)=0$ everywhere,
then from above it follows that $\Pi_{\rho\sigma}(q; D^{PT}) =0, \
q^2 \rightarrow 0$, and this is in complete agreement with
$\Pi_{\rho\sigma}(0; D^{PT})=0$, which comes from the subtraction
(3.1) at $\Delta^2(D)=0$. It worth emphasizing once more that a
massless single particle singularity appears only together with
$\Delta^2(D)$. If it is zero then there is no singularity, and the
gluon self-energy coincides with its subtracted counterpart, being
both transversal (see the initial relations (5.11) and (5.12),
which are valid only at zero $\Delta^2(D)$). The situation is
absolutely similar to the situation in QED (see subsection A in
section IV and again appendix A).

On the other hand, let us substitute the general decomposition of
the full gluon propagator (5.2) into the initial gluon SD equation
(2.1), or, equivalently, (5.3), on account of the general
relations (4.11) and (4.13), then one obtains

\begin{equation}
d(q^2) = {1 \over 1 + \Pi^s(q^2; D) + (\Delta^2(D) / q^2)},
\end{equation}
and simultaneously the relation (5.5), namely

\begin{equation}
\tilde{\Pi}(q^2; D) =0,
\end{equation}
but which requires $\Delta^2(D)=0$ as it was repeatedly discussed
above (see relations (5.5) and (5.7)). So, we need to disregard
$\Delta^2(D)$ term in the transversal component of the full gluon
propagator, shown in Eq.~(5.13), as well. In this way we are
arriving at the system of Eqs.~(5.9)-(5.10). We call this system
of equations as the PT one, since we have to put $\Delta^2(D)$
zero everywhere, and replacing $D \rightarrow D^{PT}$ in this
limit.

However, just this was the first problem. Our aim is to retain
$\Delta^2(D)$ in the gluon SD equation (5.3), or equivalently, in
the relation (5.13), and at the same time to automatically cancel
$\tilde{\Pi}(q^2; D)$ from Eq.~(5.3), i.e., without appealing to
the condition (5.14). That is why we need to introduce the
spurious mechanism which makes it possible to achieve this goal
(see especially Eq.~(6.5) below, where $\tilde{\Pi}(q^2; D)$ is
not required to be zero, i.e., it remains arbitrary). This also
makes it possible to satisfy the ST identity at non-zero
$\Delta^2(D)$, since $\tilde{\Pi}(q^2; D)$ is removed from the
gluon SD equation by the spurious technics, and therefore there is
no necessity to put it zero (see section VI).

The second problem which we have met is that at non-zero
$\Delta^2(D)$ the transversality of the full gluon self-energy
will be always violated (see the general relations (4.14) and
(4.15)). In turn this means that the ghosts cannot cancel the
longitudinal component of the full gluon propagator in this case.
This problem occurs at any gluon momentum, and not only at zero
gluon momentum $\Pi_{\rho\sigma}(0; D)= \delta_{\rho\sigma}
\Delta^2(D)$, as it follows from the subtraction (3.1) at this
point. So there is no solution of this problem at the level of the
gluon self-energy. However, we have previously noticed that, in
principle, we need rather the relevant gluon propagator to be
transversal at non-zero $\Delta^2(D)$ than the full gluon
self-energy. Just how to avoid this difficulty (and thus to
neutralize its negative consequences) for the relevant gluon
propagator at non-zero $\Delta^2(D)$, i.e., to make it transversal
without ghosts, will be explained and advocated in detail in
section IX.

\section{The general structure of the full gluon propagator}

The formal $\Delta^2(D)=0$ limit is a real way how to preserve the
color gauge invariance in QCD. Then a natural question arises why
does the mass scale parameter $\Delta^2(D)$ (which is nothing but
the re-defined tadpole term) exist in this theory at all? There is
no doubt that the color gauge invariance of QCD should be
maintained at non-zero mass scale parameter as well, since it is
explicitly present in the full gluon self-energy, and hence in the
full gluon propagator. However, by keeping it "alive", the two
important problems mentioned above arise. The first problem is how
to replace the original gluon SD equation (5.3), since it is not
consistent with the ST identity unless the mass scale parameter is
discarded from the very beginning. The second problem is how to
make the full gluon propagator purely transversal when the mass
scale parameter is explicitly present.

\subsection{The spurious mechanism }

By introducing the spurious technics here, we will be able to show
that the ST identity (5.1) can be automatically satisfied at
non-zero $\Delta^2(D)$ as well. As we already know, the original
gluon SD equation (5.3) should be correspondingly modified in this
case. In other words, our aim is to save $\Delta^2(D)$ in the
transversal invariant function (4.13), while removing it from the
longitudinal invariant function (4.12), but without going to the
formal $\Delta^2(D)=0$ limit, as it has been described in the
previous section. In order to keep the mass scale parameter
"alive", and, at the same time, to satisfy the ST identity (5.1),
it is instructive to introduce a temporary dependence on
$\Delta^2(D)$ in the free gluon propagator, thus making it an
auxiliary (spurious) free gluon propagator. The original gluon SD
equation (5.3), then should read

\begin{eqnarray}
D_{\mu\nu}(q) = D^0_{\mu\nu}(q; \Delta^2(D)) &+& D^0_{\mu\rho}(q;
\Delta^2(D))i T_{\rho\sigma}(q)[ q^2 \Pi^s(q^2; D) + \Delta^2(D)]
D_{\sigma\nu}(q) \nonumber\\
&+& D^0_{\mu\rho}(q; \Delta^2(D))i q^2 \tilde{\Pi}(q^2; D)
L_{\rho\sigma}(q) D_{\sigma\nu}(q)
\end{eqnarray}
with the spurious free gluon propagator as follows:

\begin{equation}
D^0_{\mu\nu}(q; \Delta^2(D)) =D^0_{\mu\nu}(q) +i \xi
L_{\mu\nu}(q)d_0(q^2; \Delta^2(D)) { 1 \over q^2}.
\end{equation}
We already know from the relation (5.4) that the spurious free
gluon propagator $D^0_{\mu\nu}(q; \Delta^2(D))$ can deviate from
the standard free gluon propagator (2.2) only in its longitudinal
component. The latter automatically satisfies the ST identity
(5.1), while for the former this may not be true, indeed.

Substituting this sum into the gluon SD equation (6.1), one
obtains

\begin{equation}
D_{\mu\nu}(q) = D^0_{\mu\nu}(q) + D^0_{\mu\rho}(q)i
T_{\rho\sigma}(q) [q^2 \Pi^s(q^2; D) + \Delta^2(D)]
D_{\sigma\nu}(q) + I_{\mu\nu}(q; \Delta^2(D)),
\end{equation}
where

\begin{eqnarray}
I_{\mu\nu}(q; \Delta^2(D)) &=& i \xi d_0(q^2; \Delta^2(D)) \left[
L_{\mu\nu}(q)  + L_{\mu\sigma}(q)i q^2 \tilde{\Pi}(q^2; D)
D_{\sigma\nu}(q) \right]{1 \over q^2} \nonumber\\
&+& D^0_{\mu\rho}(q)
i q^2 \tilde{\Pi}(q^2; D) L_{\rho\sigma}(q) D_{\sigma\nu}(q) \nonumber\\
&=& i \xi L_{\mu\nu}(q) \left[d_0(q^2; \Delta^2(D)) \left(1 - \xi
\tilde{\Pi}(q^2; D) \right) -\xi  \tilde{\Pi}(q^2; D) \right]{1
\over q^2}.
\end{eqnarray}
Evidently, just this term violates the ST identity (5.1) in
Eq.~(6.3), so it should be zero, which implies

\begin{equation}
d_0(q^2; \Delta^2(D)) = { \xi \tilde{\Pi} (q^2; D) \over 1 - \xi
\tilde{\Pi}(q^2; D)}.
\end{equation}
Thus the gluon SD equation (6.3) finally becomes

\begin{equation}
D_{\mu\nu}(q) = D^0_{\mu\nu}(q) + D^0_{\mu\rho}(q)i
T_{\rho\sigma}(q) [q^2 \Pi^s(q^2; D) + \Delta^2(D)]
D_{\sigma\nu}(q).
\end{equation}
Evidently, from now on we can completely forget about the spurious
free gluon propagator. It played its role and retired from the
scene.

The modified gluon SD equation (6.6) is satisfied by the same
expression for the Lorentz structure $d(q^2)$ in Eq.~(5.2) as the
original gluon SD equation (5.3) shown in Eq.~(5.13), namely

\begin{equation}
d(q^2) = {1 \over 1 + \Pi^s(q^2; D) + (\Delta^2(D) / q^2)},
\end{equation}
which is not surprising, since the original gluon SD equation
(5.3) and its modified version (6.6) differ from each other only
by the longitudinal part.

{\bf However, the important observation is that now it is not
required to put the mass scale parameter $\Delta^2(D)$ formally
zero everywhere}. The spurious mechanism does not affect the
dynamical context of the original gluon SD equation. In other
words, it makes it possible to retain the mass scale parameter in
the transversal part of the gluon SD equation, and, at the same
time, to cancel the term in its longitudinal part, which violates
the ST identity. In this way, the modified gluon SD equation (6.6)
satisfies automatically the ST identity. That is why we consider
the modified gluon SD equation (6.6) as more general than its
original counterpart (5.3).

The relation (6.7) cannot be considered as the formal solution for
the full gluon propagator $D$ in Eq.~(5.2). The general mass scale
term contribution $(\Delta^2(D) / q^2)$ and the invariant function
$\Pi^s(q^2; D)$ themselves depend on $D$. In fact, it is a
transcendental NL equation for determining $d(q^2)$ as a function
of $\Delta^2(D)$ (see section VIII below).

\section{NP QCD vs PT QCD }

In the previous section it has been explicitly shown how the gluon
SD equation should be modified in order to automatically satisfy
the ST identity at non-zero mass scale parameter. It is
instructive to collect our results here.

\subsection{NP QCD}

The modified gluon SD equation (6.6) is

\begin{equation}
D_{\mu\nu}(q; \Delta^2(D)) = D^0_{\mu\nu}(q) + D^0_{\mu\rho}(q)i
T_{\rho\sigma}(q) [q^2 \Pi^s(q^2; D) + \Delta^2(D)]
D_{\sigma\nu}(q; \Delta^2(D)),
\end{equation}
while the general tensor decomposition is the standard one (5.2),
namely

\begin{equation}
D_{\mu\nu}(q; \Delta^2(D)) = i \left\{ T_{\mu\nu}(q) d(q^2;
\Delta^2(D)) + \xi L_{\mu\nu}(q) \right\} {1 \over q^2},
\end{equation}
and the "solution" for its Lorentz structure is

\begin{equation}
d(q^2; \Delta^2(D)) = {1 \over 1 + \Pi^s(q^2; D) + (\Delta^2(D) /
q^2)}.
\end{equation}
This system of equations forms the system of equations for NP QCD,
since we distinguish between NP QCD and PT QCD by the explicit
presence of the mass scale parameter, see discussion below (that
is why we introduce temporarily the dependence on it in all the
quantities above).

\subsection{PT QCD}

The complete set of equations for PT QCD is

\begin{equation}
D^{PT}_{\mu\nu}(q) = D^0_{\mu\nu}(q) + D^0_{\mu\rho}(q)i
T_{\rho\sigma}(q) q^2 \Pi^s(q^2; D^{PT}) D^{PT}_{\sigma\nu}(q),
\end{equation}
with

\begin{equation}
D^{PT}_{\mu\nu}(q) = i \left\{ T_{\mu\nu}(q) d^{PT}(q^2) + \xi
L_{\mu\nu}(q) \right\} {1 \over q^2},
\end{equation}
and the "solution" for its Lorentz structure is

\begin{equation}
d^{PT}(q^2) = {1 \over 1 + \Pi^s(q^2; D^{PT})}.
\end{equation}

In both systems of equations the free gluon propagator is given in
Eq.~(2.2). The NP QCD system of equations has been obtained with
the help of the spurious mechanism. It made it possible to keep
the mass scale parameter $\Delta^2(D)$ "alive", and, at the same
time, to automatically satisfy the ST identity. The PT QCD system
of equations has been obtained by putting it formally zero
everywhere. Evidently, the PT system of equations can be obtained
from the NP system of equations in the formal $\Delta^2(D)=0$
limit, since the dependence of the latter system of equations on
the mass scale parameter $\Delta^2(D)$ is a regular one.

Due to asymptotic freedom (AF) in QCD the PT regime is realized at
$q^2 \rightarrow \infty$. In this limit all the Green's functions
are possible to approximate by their free PT counterparts (up to
the corresponding PT logarithms). However, from the relation (7.3)
it follows that in this limit the mass scale term contribution
$\Delta^2(D) / q^2$ is only next-to-next-to-leading order one. The
leading order contribution is the subtracted gluon self-energy
$\Pi^s(q^2; D)$, which behaves like $\ln q^2$ in this limit, as
mentioned above. The constant $1$ is the next-to-leading order
term in the $q^2 \rightarrow \infty$ limit. Such a special
structure of the relation (7.3), namely the mass scale parameter
enters it through the combination $\Delta^2(D) / q^2$ in its
denominator only, explains immediately why the mass scale
parameter $\Delta^2(D)$ is not important in PT. From this
structure it follows that the PT regime at $q^2 \rightarrow
\infty$ is effectively equivalent to the formal $\Delta^2(D)=0$ limit and vice
versa. That is the reason why this limit can be called the PT
limit. And that is why we denote $D_{\mu\nu}(q; \Delta^2=0) =
D_{\mu\nu}(q; 0) \equiv D^{PT}_{\mu\nu}(q)$, and hence $d(q^2;
\Delta^2=0) = d(q^2; 0) \equiv d^{PT}(q^2)$, etc., in accordance
with the previous notations. Let us note, however, that sometimes it is useful to distinguish
between the asymptotic suppression of the mass gap contribution $\Delta^2 / q^2$ in the
$q^2 \rightarrow \infty$ limit and the formal PT $\Delta^2=0$ limit (see our subsequent papers).

Thus the formal PT $\Delta^2(D)=0$ limit exists, and it is a
regular one. As it follows from above, in this limit one recovers
the PT QCD system of equations from the NP QCD one. So, we
distinguish between the PT and NP phases in QCD by the explicit
presence of the mass scale parameter. Its aim is to be responsible
for the NP QCD dynamics, since it dominates at $q^2 \rightarrow 0$
in the "solution" (7.3). When it is put formally zero, then the PT
phase survives only. Evidently, when such a scale is explicitly
present then the QCD coupling constant plays no role in the NP QCD
dynamics.

The mass scale parameter term does not survive in the PT $q^2
\rightarrow \infty$ regime, anyway. Then it is justified to simply
drop it in the PT. It is worth emphasizing that this does not
depend on how it has been regularized. However, as underlined
above, any regularization scheme should be adjusted to the ST
identity (5.1). In fact, in the most popular dimensional
regularization method (DRM) \cite{6} it is prescribed to put
$\Delta^2(D_0)=0$ (see also Refs. \cite{2,4} and especially the
corresponding discussion in Ref. \cite{3}). So it preserves the
color gauge invariance in PT QCD from the very beginning.

\section{The mass gap}

One of the important challenges of QCD is that the Lagrangian of
QCD \cite{1,2,3,4} does not contain a mass scale parameter which
could have a physical meaning even after the corresponding
renormalization program is performed. The only place where it
appears explicitly is the gluon SD equation of motion, as it has
been described in this work, i.e., it is only due to the
intrinsically NP (INP) dynamics of QCD developed in the gluon
sector. This underlines the importance of the investigation of the
SD system of equations and the corresponding ST identities
(\cite{1,5,7} and references therein) for understanding of the
true dynamics in the QCD ground state. The propagation of gluons
is one of the main dynamical effects there. The importance of the
corresponding equation of motion is due to the fact that its
solutions are supposed to reflect the quantum-dynamical structure
of the QCD ground state (as mentioned above, this equation is
highly NL, so the number of independent solutions is not fixed $a
\ priori$. From the very beginning they should be considered on
equal footing). The color gauge structure of this equation has
been investigated in this work.

In two-dimensional QCD the transversality condition (4.8) is
satisfied, i.e., it is zero. This means that the tadpole term
should be included from the very beginning. Otherwise, the ghosts
will not be able to cancel the longitudinal component of the full
gluon propagator \cite{2}. However, this theory has already the
scale parameter of dimension mass, which is the coupling constant.
This once more emphasizes the special status of the tadpole term,
and hence of the general mass scale parameter (4.10), in
four-dimensional QCD.

The explicit presence of the general mass scale parameter (which
is nothing but the re-defined tadpole term) in the full gluon
propagator is no coincidence. On the one hand, it does not
contradict the color gauge invariance of QCD. As it has been
explicitly shown so far in this investigation it is compatible
with the ST identity. On the other hand, it makes it possible to
introduce the mass gap so needed in NP QCD in order to explain
color confinement and other NP effects \cite{8}.

For further discussion it is convenient to re-write the relation
(7.3) in the form of the corresponding transcendental equation,
namely

\begin{equation}
d(q^2) = 1 - \left[ \Pi^s(q^2; d) + {\Delta^2(d) \over q^2}
\right] d(q^2),
\end{equation}
where instead of $D$ we introduced an equivalent dependence on
$d$, i.e., $\Delta^2(D) \equiv \Delta^2(d)$. The general mass
scale parameter $\Delta^2(d) \equiv \Delta^2(\lambda, \alpha, \xi,
g^2; d)$, where $g^2$ is the dimensionless coupling constant
squared, can be present as follows:

\begin{equation}
\Delta^2(\lambda, \alpha, \xi, g^2; d) = \Delta^2 \times
c(\lambda, \alpha, \xi, g^2; d),
\end{equation}
where the mass squared

\begin{equation}
\Delta^2 \equiv \Delta^2(\lambda, \alpha; \xi, g^2),
\end{equation}
will be called the mass gap. Contrary to the arbitrary
dimensionless constant $c(\lambda, \alpha, \xi, g^2; d)$, it does
not depend on $d$, but may, in general, depend on $\lambda,
\alpha, \xi, g^2$, and so on. Thus at this stage it is only
regularized as well as the mass scale parameter itself.

If it will survive the renormalization program, then QCD is a
complete and self-consistent theory without the need to introduce
some extra degrees of freedom in order to generate a mass gap. We
should prove that the product

\begin{equation}
\Delta^2_R = Z(\lambda, \alpha, \xi, g^2) \Delta^2(\lambda,
\alpha, \xi, g^2), \quad \lambda \rightarrow \infty, \quad \alpha
\rightarrow 0,
\end{equation}
where $Z(\lambda, \alpha, \xi, g^2)$ is the multiplicative
renormalization (MR) constant of the mass gap, exists in the
above-shown limits. However, the final result of these limits,
i.e., $\Delta^2_R$, should be achieved in the way not to
compromise the general renormalizability of QCD. Contrary to the
regularized version, the renormalized mass gap (8.4) should not
depend on the gauge-fixing parameter, should be finite, positive
definite, etc. Only after performing this program, one can assign
to it a physical meaning of a scale responsible for the true NP
dynamics of QCD  at large distances, in the same way as
$\Lambda^2_{QCD}$ is responsible for its nontrivial PT dynamics at
short distances. Apparently, the renormalized mass gap can be
identified/related with/to the Jaffe and Witten mass gap \cite{8},
if it is possible at all. The MR program will be positively
resolved in the subsequent papers. For this we have to explicitly
find $d(q^2)$ as a function of the regularized mass gap $\Delta^2$
in the general way, and in particular with the help of Eq. (8.1).

Concluding, it is worth emphasizing that the formal PT
$\Delta^2(d)=0$ limit implies to put the mass gap formally zero as
well, namely $\Delta^2=0$. The arbitrary coefficient $c(\lambda,
\alpha, \xi, g^2; d)$, which appears in Eq.~(8.2), is, in general,
not zero. In the rest of this paper instead of the general mass
scale parameter $\Delta^2(d)$ we will use the mass gap $\Delta^2$
itself, for simplicity.

\section{Restoration of transversality of the gluon propagator in NP QCD}

The NP QCD system of equations(7.1)-(7.3) depends explicitly on
the mass gap $\Delta^2$. As we already know from above, then the
ghosts are not able to cancel the longitudinal component in the
full gluon propagator (7.2), i.e., they are of no use in this case
(the transversality condition for the full gluon self-energy is
always violated, see relations (4.14) and (4.15)). This is the
price we have paid to keep the mass gap "alive" in the full gluon
propagator. Our aim here is to formulate a method which allows one
to make the gluon propagator, relevant for NP QCD, purely
transversal in a gauge invariant way, even if the mass gap is
explicitly present.

For this purpose let us define the truly NP (TNP) part of the full
gluon propagator as follows:

\begin{equation}
D^{TNP}_{\mu\nu}(q; \Delta^2) = D_{\mu\nu}(q; \Delta^2) -
D_{\mu\nu}(q; \Delta^2=0) = D_{\mu\nu}(q; \Delta^2) -
D^{PT}_{\mu\nu}(q),
\end{equation}
i.e., the subtraction is made with respect to the mass gap
$\Delta^2$, and therefore the separation between these two terms
is exact. Evidently, the formal PT $\Delta^2(D)=0$ limit is
equivalently replaced by the formal mass gap limit to zero, i.e.,
$\Delta^2=0$, as underlined above. So on account of the
expressions (7.2) and (7.5), it becomes

\begin{equation}
D^{TNP}_{\mu\nu}(q; \Delta^2) = i T_{\mu\nu} (q) \Bigr[ d(q^2;
\Delta^2) -  d^{PT}(q^2) \Bigl] {1 \over q^2} = i T_{\mu\nu} (q)
d^{TNP}(q^2; \Delta^2) {1 \over q^2},
\end{equation}
where the explicit expression for the TNP Lorentz structure
$d^{TNP}(q^2; \Delta^2) = d(q^2;
\Delta^2) -  d^{PT}(q^2)$ can be obtained from the relations (7.3)
and (7.6) for $d(q^2; \Delta^2)$ and $d^{PT}(q^2)$, respectively.

The subtraction (9.1) is equivalent to

\begin{equation}
D_{\mu\nu}(q; \Delta^2) = D^{TNP}_{\mu\nu}(q; \Delta^2) +
D^{PT}_{\mu\nu}(q).
\end{equation}
The TNP gluon propagator (9.2) does not survive in the formal PT
$\Delta^2=0$ limit. This means that it is free of the PT
contributions, by construction. The full gluon propagator (7.2) in
this limit is reduced to its PT counterpart (7.5). This means that
the full gluon propagator, being also NP, nevertheless, is
"contaminated" by them. The TNP gluon propagator is purely
transversal in a gauge invariant way (no special (Landau) gauge
choice by hand), while its full counterpart has a longitudinal
component as well. There is no doubt that the true NP dynamics of
the full gluon propagator is completely contained in its TNP part,
since the subtraction (9.3) is nothing but adding zero to the full
gluon propagator. We can write

\begin{eqnarray}
D_{\mu\nu}(q; \Delta^2) &=& i \left\{ T_{\mu\nu}(q) d(q^2;
\Delta^2) + \xi L_{\mu\nu}(q) \right\} {1 \over q^2} - i
T_{\mu\nu}(q) d^{PT}(q^2){1 \over q^2} + i T_{\mu\nu}(q)
d^{PT}(q^2){1 \over q^2} \nonumber\\
&=& D^{TNP}_{\mu\nu}(q; \Delta^2) + D^{PT}_{\mu\nu}(q),
\end{eqnarray}
and so the true NP dynamics in the full gluon propagator is not
affected, but contrary exactly separated from its PT dynamics,
indeed. In other words, the TNP gluon propagator is the full gluon
propagator but free of its PT "tail".

\subsection{Prescription}

Taking this important observation into account, we propose instead
of the full gluon propagator (7.2) to use its TNP counterpart
(9.2) as the relevant gluon propagator for NP QCD, i.e., to
replace

\begin{equation}
D_{\mu\nu}(q; \Delta^2)  \rightarrow D^{TNP}_{\mu\nu}(q; \Delta^2)
= D_{\mu\nu}(q; \Delta^2) - D^{PT}_{\mu\nu}(q),
\end{equation}
and hence

\begin{equation}
d(q^2; \Delta^2) \rightarrow d^{TNP}(q^2; \Delta^2) = d(q^2;
\Delta^2) - d^{PT}(q^2).
\end{equation}
The subtraction (9.5) plays effectively the role of ghosts in our
proposal (for its additional arguments and motivation see appendix
B). However, the ghosts cancel only the longitudinal component in
the PT gluon propagator (7.5), while our proposal leads to the
cancellation of the PT contribution in the full gluon propagator
(7.2) as well (and thus to an automatical cancellation of its
longitudinal component). Nevertheless, this is not a problem,
since the mass gap is not survived in the formal PT limit, anyway.

In fact, our proposal is reduced to a rather simple prescription.
If one knows a full gluon propagator, and is able to identify the
mass scale parameter responsible for the NP dynamics in it, then
the full gluon propagator should be replaced in accordance with
the subtraction (9.5). The only problem with it is that, being
exact, it may not be unique. However, the uniqueness of such kind
of separation can be achieved only in the explicit solution for
the full gluon propagator as a function of the mass gap (see the
above-mentioned subsequent papers). Anyway, this subtraction is a
first necessary step, which guarantees transversality of the TNP
gluon propagator $D^{TNP}_{\mu\nu}(q; \Delta^2)$ without losing
even one bit of information on the true NP dynamics in the full
gluon propagator $D_{\mu\nu}(q; \Delta^2)$. At the same time, its
non-trivial PT dynamics is completely saved in its PT part
$D^{PT}_{\mu\nu}(q)$. So it is worth emphasizing that the both
terms in the subtraction (9.3) are valid in the whole momentum
range, i.e., they are not asymptotics.

The full gluon propagator (7.2), keeping the mass gap "alive", is
not "physical" in the sense that it cannot be made transversal by
ghosts. Therefore it cannot be used for numerical calculations of
the physical observables from first principles.  However, our
proposal makes it possible to present it as the exact sum of the
two "physical" propagators. The TNP gluon propagator is
automatically transversal, by construction. It fully contains all
the information of the full gluon propagator on its NP context.
Just it should be used in accordance with the prescription (9.5)
in order to calculate the physical observables in low-energy QCD.

On the other hand, in high-energy QCD the full gluon propagator
should be replaced as follows:

\begin{equation}
D_{\mu\nu}(q; \Delta^2)  \rightarrow D^{PT}_{\mu\nu}(q) =
D_{\mu\nu}(q; \Delta^2) - D^{TNP}_{\mu\nu}(q; \Delta^2),
\end{equation}
and hence

\begin{equation}
d(q^2; \Delta^2) \rightarrow d^{PT}(q^2) = d(q^2; \Delta^2) -
d^{TNP}(q^2; \Delta^2).
\end{equation}
The PT gluon propagator $D^{PT}_{\mu\nu}(q)$ is free of the mass
gap, and hence the ghosts will cancel its longitudinal component
in accordance with the transversality relation (4.9), making it
thus transversal ("physical"). The PT gluon propagator
$D^{PT}_{\mu\nu}(q)$ fully contains all the information of the
full gluon propagator on its non-trivial PT context (scale
violation, AF \cite{1,2,3,4}).

It is instructive to obtain Eq.~(9.2) by the substitution into the
subtraction (9.1) not the corresponding "solutions" but rather the
corresponding modified gluon SD equation (7.1) itself and
Eq.~(7.4). Doing so, one obtains

\begin{eqnarray}
D^{TNP}_{\mu\nu}(q; \Delta^2) &=& D^0_{\mu\rho}(q)i
T_{\rho\sigma}(q) \left[ q^2 \Pi^s(q^2; D) + \Delta^2 \right]
D_{\sigma\nu}(q; \Delta^2) \nonumber\\
&-& D^0_{\mu\rho}(q)i T_{\rho\sigma}(q) q^2 \Pi^s(q^2; D^{PT})
D^{PT}_{\sigma\nu}(q).
\end{eqnarray}
Substituting further into this equation the expressions (2.2),
(7.2) and (7.5) and after doing some tedious algebra, one arrives
at

\begin{equation}
D^{TNP}_{\mu\nu}(q; \Delta^2) = - i T_{\mu\nu}(q) \left[
\Pi^s(q^2; D) + {\Delta^2 \over q^2} \right] d(q^2; \Delta^2) {1
\over q^2} + i T_{\mu\nu}(q) \Pi^s(q^2; D^{PT}) d^{PT}(q^2) {1
\over q^2}.
\end{equation}
From the relations (7.3) and (7.6) it follows that

\begin{equation}
\left[ \Pi^s(q^2; D) + {\Delta^2 \over q^2} \right] d(q^2;
\Delta^2) = 1 - d(q^2; \Delta^2), \quad \Pi^s(q^2; D^{PT})
d^{PT}(q^2)  = 1 - d^{PT}(q^2),
\end{equation}
and substituting them back into the previous equation, one again
arrives at Eq.~(9.2).

Let us show explicitly the SD equation for the TNP gluon
propagator. From Eq.~(9.9), on account of the subtraction (9.3),
one gts

\begin{eqnarray}
D^{TNP}_{\mu\nu}(q; \Delta^2) &=& D^0_{\mu\rho}(q)i
T_{\rho\sigma}(q)[ - q^2 \Pi^s(q^2; D^{PT}) + q^2 \Pi^s(q^2; D)
+ \Delta^2] D^{PT}_{\sigma\nu}(q) \nonumber\\
&+& D^0_{\mu\rho}(q)i T_{\rho\sigma}(q) [q^2 \Pi^s(q^2; D) +
\Delta^2] D^{TNP}_{\sigma\nu}(q; \Delta^2),
\end{eqnarray}
where $D^{PT}_{\sigma\nu}(q)$ satisfies its own Eq.~(7.4).

The remarkable feature of this equation is that, by switching
interaction off (i.e., setting formally $\Pi^s(q^2; D^{PT}) =
\Pi^s(q^2; D) = \Delta^2 =0$), it cannot be reduced to the free
gluon propagator, like this occurs for the full gluon propagator
(7.1) and its PT counterpart (7.4). In other words, in the truly
NP QCD the gluon propagator is always "dressed". And this brings
in one more serious argument in favor of the above proposed
subtractions of all the types of the PT contributions. The
emission and absorbtion of the colored dressed gluons at large distances can be
suppressed by the renormalization of the mass
gap (see subsequent papers). At the same time, there exists no
such mechanism to do the same with the colored free gluons in
order to ensure their confinement. So the correct theory of
low-energy QCD should exclude the free gluon propagator from its
formalism. This just takes place in the truly NP QCD as a result
of the subtraction of the PT gluon propagator, which always
contains the free gluon propagator.

Concluding, the solution of the above-mentioned two problems how
to preserve the color gauge invariance/symmetry in QCD at non-zero
mass gap completes our investigation in this paper. This means
that from now on we can forget the relations (4.14) and (4.15) at
all, since there are no any more their negative consequences for
the truly NP QCD. In this connection let us underline that the
initial subtraction (3.1) has been done in a gauge invariant way
(i.e., not in separate propagators, which enter the skeleton loop
integrals, contributing to the full gluon self-energy).

\section{General discussion}

The general scale parameter (4.10), having the dimensions of mass
squared, or, equivalently, the mass gap $\Delta^2$ is dynamically
generated in the QCD gluon sector. It is mainly due to the
non-linear interaction of massless gluon modes. It is defined as
the value of the full gluon self-energy at some finite point. Thus
it has not been introduced by hand, since it is hidden in the
skeleton loop integrals, contributing to the full gluon
self-energy. To make its existence perfectly clear just the
definition of the subtracted gluon self-energy in Eq.~(3.1) has
been proposed.

As pointed out above, the general scale parameter (4.10) is
nothing but the re-defined constant skeleton tadpole term.
Moreover, it is reduced to the tadpole term itself if one puts in
the relation (4.10) $\Delta^2_g(D)=0$ in accordance with the
transversality condition (4.9). So the tadpole term

\begin{equation}
\Pi_t(D) = \Delta^2_t(D) = g^2 \int {i d^4 q_1 \over (2 \pi)^4}
T^0_4 (q_1, 0,0, -q_1)D(q_1)
\end{equation}
plays a key role in the dynamical generation of the mass gap (for
simplicity, we omit the tensor and color indices). In its turn, it
is explicitly generated by the point-like four-gluon vertex only.
The triple gluon vertex vanishes when all the gluon momenta
involved go to zero ($T_3(0,0) =0$), while its four-gluon
counterpart survives ($T_4(0,0,0) \neq 0$). In this connection let
us remind that the mass gap dominates the structure of the gluon
SD equation (7.1) and its "solution" (7.3) just in this limit. So
there is no doubt in the important role of the quartic gluon
vertex in NP QCD. At the same time, in the dynamical generation of
the re-defined tadpole term (4.10) all the QCD full gluon vertices
are explicitly involved.

All the quantities considered in this paper are necessarily
regularized, as a first step. However, nothing depends in our
approach on the specific regularization scheme, preserving or not
gauge invariance. It is impossible to perform any concrete
calculations of the regularized skeleton loop integrals,
containing unknown, in general, the full propagators and vertices.
No any truncations/approximations/assumptions (which means no use
of PT), special gauge choice, etc., have been made for them. Only
algebraic, i.e., exact derivations have been done in this paper.

The mass gap $\Delta^2$ violates explicitly the ST identity for
the full gluon propagator, which satisfies the corresponding
equation of motion. Also, in its presence the ghosts are not able
to cancel the longitudinal component in the full gluon propagator
in order to guarantee unitarity of the $S$-matrix in this theory.
So it should be disregarded on the general grounds, i.e., put
formally zero everywhere. We have explicitly shown that this
formal limit is equivalent to the PT $q^2 \rightarrow \infty$
limit and vice versa, leading thus to the formulation of the
system of equations for PT QCD.

In order to confirm that the color gauge invariance/symmetry of
QCD is maintained in the explicit presence of the mass gap, we
have introduced the spurious mechanism. It makes it possible to
modify the original gluon SD equation for the full gluon
propagator in a such way that makes the ST identity (5.1)
automatically satisfied at non-zero $\Delta^2$. At the same time,
the dynamical context of the modified gluon SD equation is not
affected, i.e., it is the same as of the original gluon SD
equation. The "solution" (7.3) depends regularly on the mass gap,
and it has a correct PT $\Delta^2=0$ limit, shown in the relation
(7.6). From it clearly follows that the effect of the mass gap
dominates at $q^2 \rightarrow 0$ and strongly suppressed in the PT
$q^2 \rightarrow \infty$ regime, so it is justified to simply
disregard it in PT.

The standard gluon SD equation (2.1) suffers from the overlapping
UV divergences. Its counterpart, which is free of them \cite{9}
(and references therein), has a much more complicated structure
than Eq.~(2.1). However, we hope that using the same technics (or
its a more sophisticated version) one can achieve the same
conclusion, that's the mass gap is consistent with the color gauge
invarince/symmetry. Quite possible that there is no point in this
investigation, since the mass gap does not survive in the PT $q^2
\rightarrow \infty$ limit (see a brief discussion in Ref.
\cite{10} as well). Anyway, this investigation should be done
elsewhere.

Our "solution" (7.3) depends explicitly on the mass gap. As
underlined above, in this case the ghosts are of no use to cancel
the longitudinal component in the full gluon propagator. However,
we have formulated a general method which makes it possible to
achieve transversality of the full gluon propagator, relevant for
NP QCD, in a gauge invariant way. It is based on the exact
subtraction of the PT contribution from the full gluon propagator.
Such obtained the TNP gluon propagator is purely transversal,
maintaining thus unitarity of the $S$-matrix within our approach.
It completely reproduces the true NP structure of the full gluon
propagator, and, at the same time, is free of the PT
"contaminations" at the fundamental gluon level.

As pointed out above, we need no ghosts to ensure the cancellation
of the longitudinal component in the full gluon propagator.
Nevertheless, this does not mean that we need no ghosts at all. We
need them in other sectors of QCD, for example in the quark ST
identity, which contains the so-called ghost-quark scattering
kernel explicitly \cite{1}. This kernel still makes an important
contribution to the identity even if the relevant gluon propagator
is transversal \cite{11,12,13,14,15}. Do not mix the TNP gluon
propagator (9.2) with the full gluon propagator (7.2) in the
Landau gauge. The former is transversal by construction in a gauge
invariant way. The latter one becomes transversal only by choosing
Landau gauge by hand, i.e., not in a gauge invariant way (let us
remind that the ghosts cannot make it transversal).

In place of mass gap, any other mass scale parameter might serve.
This could be introduced into the full gluon propagator by hand,
as an ansatz, or arise as a result of some specific
approximation/truncation made in the gluon SD equation itself and
hence in its solution, etc. Its origin is irrelevant for our
method, the only request is that the full gluon propagator should
regularly depend on it. However, none of the
truncations/approximations or ansatzs made or introduced in the
framework of any approach should undermine the above-discussed
general role of ghosts in PT QCD. Our method just guarantees this.

\section{Conclusions}

We have discussed some important issues of the color gauge
invariance/symmetry of QCD without use of PT. The basic relation
of our analysis is the subtraction (3.1), which clearly shows the
NL dynamical origin of the mass gap in the gluon sector of QCD. It
has the dimensions of mass squared. All the quantities are
necessary regularized, and only algebraic derivations have been
done with them. The mass gap violates the ST identity (5.1) for
the full gluon propagator, which satisfies the corresponding
equation of motion. It also prevents the ghosts to cancel the
longitudinal component in the full gluon propagator. In order to
maintain the color gauge invariance/symmetry it should be
disregarded from the very beginning, i.e., put formally zero
everywhere.

However, by introducing the initial subtraction (3.1) and the
spurious technics, we have explicitly shown how to satisfy the ST
identity at non-zero mass gap as well. Our approach makes it
possible to retain it in the transversal part of the gluon SD
equation, while cancelling the term in its longitudinal part,
which violates the ST identity. So the modified gluon SD equation
automatically satisfies the ST identity for the full gluon
propagator. At the same time, its dynamical context is not
affected. Its "solution" (7.3) depends regularly on the mass gap
term, and has a correct PT limit, i.e., the mass gap contribution
does not survive in the PT $q^2 \rightarrow \infty$ regime. At the
same time, the mass gap contribution dominates the structure of
the "solution" (7.3) in the NP $q^2 \rightarrow 0$ limit.

We have also formulated a general method which allows to derive
the gluon propagator relevant for NP QCD, the so-called TNP gluon
propagator (9.2). It regularly depends on the mass gap in the way
when it is put formally zero then it also vanishes. The basic
element in this method is the subtraction (9.1). It exactly
separates the TNP part from its PT counterpart in the full gluon
propagator. The TNP gluon propagator is purely transversal in a
gauge invariant way, maintaining thus unitarity of the $S$-matrix
in NP QCD. It completely reproduces the true NP structure of the
full gluon propagator, and, at the same time, is free of the PT
contributions ("contaminations"). Thus it cannot be reduced to the
free gluon propagator when the interaction is formally switched
off. This is important to correctly understand the confinement
mechanism in QCD. As emphasized above, the emission and absorbtion
of free gluons at large distances cannot be suppressed, in principle. The TNP QCD
provides the solution of this fundamental problem, and therefore
the proposed subtraction (9.1) seems to be necessary. It makes the
relevant gluon propagator transversal and excludes the free gluons
from the theory at the same time. That is why we have argued that
just the TNP QCD should be used in order to correctly calculate
the physical observables, processes, etc. in low-energy QCD.

Briefly, our approach to NP QCD is based on the initial definition
(3.1) and the above-mentioned spurious (section VI) and
subtraction (section IX) methods. It makes it possible to maintain
the ST identity for the full gluon propagator, when the mass gap
(or any other mass scale parameter) is explicitly present in it.
It also makes it possible to neutralize the negative consequences
of the violation of the transversality condition for the full
gluon self-energy, when the mass gap is kept "alive". All other ST
identities are not affected, and the color currents are conserved
as well as no particular gauge choice made.

Concluding, we arrived at a dilemma. To maintain the color gauge
symmetry of QCD and unitarity of its $S$-matrix the mass gap
should be dropped, coming thus to PT QCD. To keep the mass gap
then one should proceed in accordance with our method. It allows
one to maintain both the above-mentioned symmetry and unitarity in
a gauge invariant way, coming thus to the truly NP QCD. We
distinguish between them by the explicit presence of the mass gap
and not by the strength of the coupling constant. It plays no role
when the mass gap is kept "alive". On the other hand, the
difference between the truly NP and NP QCD is that the former
vanishes in the PT $\Delta^2=0$ limit, while the latter survives,
and is to be reduced to PT QCD (we mean the corresponding
equations of motion (7.1), (9.12) and (7.4), of course).

{\bf The common belief (which comes from PT) that the mass gap
(which is nothing but the re-defined tadpole term) contradicts the
color gauge invariance/symmetry of QCD is false. This fundamental
symmetry is maintained/preserved at non-zero mass gap as well}.

\begin{acknowledgments}

Support by HAS-JINR grant (P. Levai) is to be
acknowledged. The author is grateful to P. Forg\'{a}cs, J. Nyiri, V. Skokov, M.
Faber, C. Wilkin, T. Bir\'{o}, \'{A}. Luk\'{a}cs, M. Vas\'{u}th and to
A.V. Kouzushin for useful discussions, remarks and help.

\end{acknowledgments}

\appendix

\section{QED}

It is instructive to discuss in more detail (than it has been done
in subsection A of section IV) why the mass gap does not occur in
QED. The photon SD equation for the full photon propagator $D(q)$
can be symbolically written down as follows:

\begin{equation}
D(q) = D_0(q) + D_0(q) \Pi(q) D(q),
\end{equation}
where we omit, for convenience, the dependence on the Dirac
indices, and $D_0 \equiv D_0(q)$ is the free photon propagator.
$\Pi(q)$ describes the vacuum polarization tensor contribution to the
photon self-energy. Analytically it looks

\begin{equation}
\Pi(q) \equiv \Pi_{\mu\nu}(q) = - g^2 \int {i d^4 p \over (2
\pi)^4} Tr [\gamma_{\mu} S(p-q) \Gamma_{\nu}(p-q, q)S(p)],
\end{equation}
where $S(p)$ and $\Gamma_{\mu}(p-q,q)$ represent the full electron
propagator and the full electron-photon vertex, respectively. Here
and everywhere below the signature is Euclidean, since it implies
$q_i \rightarrow 0$ when $q^2 \rightarrow 0$ and vice-versa. This
tensor has the dimensions of mass squared, and therefore it is
quadratically divergent. It should be regularized (see remarks
below).

Similar to the QCD case, let us introduce the mass gap through the
definition of the subtracted photon self-energy as follows:

\begin{equation}
\Pi^s(q) \equiv \Pi^s_{\mu\nu}(q) = \Pi_{\mu\nu}(q) -
\Pi_{\mu\nu}(0) = \Pi_{\mu\nu}(q) - \delta_{\mu\nu}\Delta^2
(\lambda),
\end{equation}
where the dimensionless UV regulating parameter $\lambda$ has been
introduced into the mass gap $\Delta^2(\lambda)$, given by the
integral (A2) at $q^2=0$, in order to assign a mathematical
meaning to it. In this connection let us note that what we have
said about the regularization of all the quantities in section III
is valid here as well, apart from one observation. The
above-mentioned subtraction at zero point $q^2=0$ is not dangerous
in QED, since it is an abelian quantum gauge theory. In this
theory there is no the self-interaction of massless photons, which
may be source of the singularities in the $q^2 \rightarrow 0$
limit.

The decompositions of the vacuum polarization tensor and its
subtracted counterpart into the independent tensor structures can
be written as follows:

\begin{eqnarray}
\Pi_{\mu\nu}(q) = T_{\mu\nu}(q) q^2 \Pi_1(q^2) + q_{\mu}
q_{\nu}(q) \Pi_2(q^2), \nonumber\\
\Pi^s_{\mu\nu}(q) = T_{\mu\nu}(q) q^2 \Pi^s_1(q^2) + q_{\mu}
q_{\nu}(q) \Pi^s_2(q^2),
\end{eqnarray}
where all the invariant functions of $q^2$ are dimensionless ones.
In addition, $\Pi^s_n(q^2)$ at $n=1,2$  cannot have the pole-type
singularities in the $q^2 \rightarrow 0$ limit, since $\Pi^s(0)
=0$, by definition; otherwise all the invariant functions remain
arbitrary. From these relations it follows that $\Pi^s(q) =
O(q^2)$, i.e., it is always of the order $q^2$.

Substituting these decompositions into the subtraction (A3), one
obtains

\begin{eqnarray}
\Pi_2(q^2) = \Pi^s_2(q^2) + {\Delta^2(\lambda)
\over q^2}, \nonumber\\
\Pi_1(q^2) = \Pi^s_1(q^2) + {\Delta^2(\lambda) \over q^2}.
\end{eqnarray}

On the other hand from the transversality condition for the photon
self-energy

\begin{equation}
q_{\mu} \Pi_{\mu\nu}(q) =q_{\nu} \Pi_{\mu\nu}(q) =0,
\end{equation}
which comes from the current conservation in QED, one arrives at
$\Pi_2(q^2)=0$. Then from the relations (A5) it follows that

\begin{equation}
\Pi^s_2(q^2)= - {\Delta^2(\lambda) \over q^2},
\end{equation}
which, however, is impossible since $\Pi^s_2(q^2)$ cannot have a
massless single particle singularity, as mentioned above. So the
mass gap should be discarded, i.e., put formally zero and,
consequently, $\Pi^s_2(q^2)$ as well, i.e.,

\begin{equation}
\Delta^2(\lambda)=0, \quad \Pi^s_2(q^2)=0.
\end{equation}
Thus the subtracted photon self-energy is also transversal, i.e.,
satisfies the transversality condition $q_{\mu} \Pi^s_{\mu\nu}(q)
=q_{\nu} \Pi^s_{\mu\nu}(q) =0$. This means that it coincides with
the photon self-energy. This comes from the subtraction (A3), on
account of the relations (A8), i.e.,

\begin{equation}
\Pi_{\mu\nu}(q) = \Pi^s_{\mu\nu}(q) = T_{\mu\nu}(q) q^2
\Pi^s_1(q^2),
\end{equation}
so that the photon self-energy does not have a pole in its
invariant function $\Pi_1(q^2)= \Pi^s_1(q^2)$. In obtaining these
results neither the PT has been used nor a special gauge has been
chosen. So there is no place for quadratically divergent constant
in QED, while logarithmic divergence still can be present in the
invariant function $\Pi_1(q^2) = \Pi^s_1(q^2)$. It is to be
included into the electric charge through the corresponding
renormalization program (for these detailed gauge-invariant
derivations explicitly done in lower order of the PT see any
text-book on QED).

In fact, the current conservation condition (A6), i.e.,
transversality of the photon self-energy lowers the quadratic
divergence of the corresponding integral (A2) to a logarithmic
one. That is the reason why in QED only logarithmic divergences
survive. The current conservation condition for the photon
self-energy (A6) and the ST identity for the full photon
propagator $q_{\mu}q_{\nu}D_{\mu\nu}(q) = i\xi$ are consequences
of gauge invariance. They should be maintained at every stage of
the calculations, since the photon is a physical state. In other
words, at all stages the current conservation plays a crucial role
in extracting physical information from the $S$-matrix elements in
QED, which are usually proportional to the combination $j^{\mu}_1
(q)D_{\mu\nu}(q) j^{\nu}_2(q)$. The current conservation condition
$j^{\mu}_1 (q) q_{\mu} = j^{\nu}_2(q)q_{\nu} =0$ implies that the
unphysical (longitudinal) component of the full photon propagator
does not change the physics of QED, i.e., only its physical
(transversal) component is important. In its turn this means that
the transversality condition imposed on the photon self-energy is
also important, because $\Pi_{\mu\nu}(q)$ itself is a correction
to the amplitude of the physical process, for example such as
electron-electron scattering.

Thus in QED there is no mass gap, and we can replace $\Pi(q)$ by
its subtracted counterpart $\Pi^s(q)$ from the very beginning
($\Pi(q) \rightarrow \Pi^s(q)$), totally discarding the
quadratically divergent constant $\Delta^2(\lambda)$ from all the
equations and relations. Then the photon SD equation (A1) becomes

\begin{equation}
D(q) = D_0(q) + D_0(q) \Pi^s(q) D(q),
\end{equation}
which can be summed up into geometric series, so one has

\begin{equation}
D(q) = {D_0(q) \over  1 - \Pi^s(q) D_0(q)}= D_0(q) + D_0(q)
\Pi^s(q) D_0(q) - D_0(q) \Pi^s(q)D_0(q) \Pi^s(q)D_0(q) + ... \ .
\end{equation}
Since $\Pi^s(q) = O(q^2)$ and $D_0(q) \sim (q^2)^{-1}$, the IR
singularity of the full photon propagator is determined by the IR
singularity of the free photon propagator, which is always $1 /
q^2$. Hence the photons even "dressed" always remain massless (for the description of QED see any corresponding textbook but especially Ref. \cite{16}). 

Concluding, we cannot release the mass gap from the QED vacuum,
while we can release the photons and the electron-positron pars
from it. In QCD the situation is completely opposite to QED. In
this theory we can release the mass gap from its vacuum, as it has
been described in this work. But we cannot release the gluons and
the quarks/antiquarks from the QCD vacuum because of the color
confinement phenomenon. When we speak about the mass gap to be released or not from the
vacua of the corresponding quantum gauge theories, we mean, of course, should it not be discarded
or to be discarded from the very beginning in these theories. In this connection, let us remind that
at this stage the mass gap is not physical, it is only regularized quantity.

\section{Motivation}

An ultimate goal of any fundamental theory like QCD is to describe
the physical observables, processes, etc., from first principles.
It has been already achieved in PT QCD, which correctly describes
the behavior of QCD in the high energy/momentum limit. To do the
same in the low-energy/momentum region is a formidable task
because of the color confinement phenomenon, the dynamical
mechanism of which is not yet understood \cite{17,18,19,20}. However,
first what we need in order to accomplish the above-mentioned goal
in low-energy QCD is to define correctly the relevant gluon
propagator; it should be purely transversal and should reproduce
the true large-scale structure of the QCD ground state.

In order to justify our proposal how to satisfy these conditions,
let us discuss briefly one of the important characteristics of the
QCD ground state -- the Bag constant. It is just defined as the
difference between the PT and NP vacuum energy densities (VED)
\cite{21,22,23,24}. So, we can symbolically put $B = VED^{PT} - VED$,
where $VED$ is the NP but "contaminated" by the PT contributions
(i.e., this is a full $VED$ like the full gluon propagator). At
the same time, in accordance with our method we can continue as
follows: $B = VED^{PT} - VED = VED^{PT} - [VED - VED^{PT} +
VED^{PT}] = VED^{PT} - [VED^{TNP} + VED^{PT}] = - VED^{TNP} > 0$,
since the VED is always negative. Thus the Bag constant is nothing
but the TNP VED, apart from the sign, by definition, and thus is
completely free of the PT "contaminations".

An adequate formalism for the calculation of the Bag constant from
first principles is the effective potential approach for composite
operators \cite{25,26}. In the absence of external sources it is
nothing but the VED. To leading order it gives the VED as a
special function of the Lorentz structure in the full gluon
propagator, which should be then integrated out over its momentum.
Thus, in order to correctly calculate the Bag constant in
accordance with the above-mentioned definition, it is necessary to
replace the full Lorentz structure by its TNP counterpart due to
the prescription (9.6) within our method. For how to correctly
define and actually calculate the Bag constant from first
principles, by making all the necessary subtractions at all
levels, see Ref. \cite{27,28}.

In turn, via the well known relations (see remarks below) the Bag
constant is related to many other important NP quantities in QCD,
such as the gluon and quark condensates, the topological
susceptibility, etc., which are defined beyond PT only
\cite{21,22,28,29}. This means that they are determined by such
correlation functions from which all the types of the PT
contributions should be, by definition, subtracted. It is worth
emphasizing that such type of the subtractions are inevitable also
for the sake of self-consistency. As mentioned above, in
low-energy QCD there exist relations between different correlation
functions, for example, the Witten-Veneziano (WV) and
Gell-Mann-Oakes-Renner (GMOR) formulae. The former \cite{30,31,32,33,34}
relates the pion decay constant and the mass of the $\eta'$ meson
to the topological susceptibility. The latter \cite{21,22,35} relates
the chiral quark condensate to the pion decay constant and its
mass. The famous trace anomaly relation (see, for example Refs.
\cite{21,22,32,33,34} and references therein) relates the Bag constant
to the gluon and quark condensates \cite{21,22,27}. Defining thus the
Bag constant, the topological susceptibility and the gluon and
quark condensates by the subtraction of all the types of the PT
contributions, it would not be self-consistent to retain them in
the correlation function, determining the pion decay constant, and
in the expressions for the pion and $\eta'$ meson masses.

A few additional remarks about the subtraction of the PT
contributions are in order. Let us remind that in lattice QCD
\cite{36} such kind of an equivalent procedure also exists. In
order to prepare an ensemble of lattice configurations for the
calculation of any NP quantity or to investigate some NP
phenomena, the excitations and fluctuations of gluon fields of the
PT origin and magnitude should be "washed out" from the vacuum.
This goal is usually achieved by using "Perfect Actions",
"cooling", "cycling", etc., (see Refs. \cite{17,18} and references
therein). Evidently, this is rather similar to our method in
continuous QCD.

In QCD sum rules approaching the deep IR region from above, the IR
sensitive contributions were parameterized in terms of a few
quantities (the gluon and quark condensates, etc.), while the
direct access to NP effects was blocked by the IR divergences
\cite{21,22,37}. In the phenomenological estimate of the gluon
condensate by this method the PT gluon propagator integrated out
over the deep IR region (where it certainly fails) is to be
dropped (see discussion given by Shifman in Ref. \cite{18}. To
drop the term in one side of the equation, relation, etc. is
equivalent to subtract the same term from its another side). The
necessity of the subtraction of the PT part of the "running"
coupling constant (integrated out) for the analytical calculation
of the gluon condensate has been explicitly shown in Ref.
\cite{38}.

In order to calculate correctly any truly NP quantity in
low-energy QCD from first principles one has to begin with making
subtractions at the fundamental quark-gluon level, anyway. The
necessity of the subtractions at all levels is discussed in Ref.
\cite{10} in more detail. Our proposal (9.5)-(9.6) ensures
transversality of the TNP gluon propagator in a gauge invariant
way. This makes it possible to maintain unitarity of the
$S$-matrix, even if the mass gap (or any other mass scale
parameter) is explicitly present. The TNP gluon propagator
correctly reproduces the true NP structure of the full gluon
propagator, while being free of its PT contributions. This also
excludes the free gluons from the theory, which, as underlined
above, is important for the correct understanding of the color
confinement mechanism.

\end{document}